\documentclass[]{interact}

\usepackage{epstopdf}
\usepackage[caption=false]{subfig}

\usepackage[longnamesfirst,sort]{natbib}
\bibpunct[, ]{(}{)}{;}{a}{,}{,}
\renewcommand\bibfont{\fontsize{10}{12}\selectfont}

\theoremstyle{plain}

\theoremstyle{definition}

\theoremstyle{remark}

\begin{document}


\title{Automatic end-to-end De-identification: Is high accuracy the only metric?}

\author{
\name{Vithya Yogarajan\textsuperscript{a}\thanks{CONTACT Vithya Yogarajan. Email: vyogaraj@waikato.ac.nz},  Bernhard Pfahringer\textsuperscript{a} and Michael Mayo\textsuperscript{a}}
\affil{\textsuperscript{a}Department of Computer Science, The University of Waikato, Hamilton, New Zealand}
}

\maketitle

\begin{abstract}
De-identification of electronic health records (EHR) is a vital step towards advancing health informatics research and maximising the use of available data. It is a two-step process where step one is the identification of protected health information (PHI), and step two is replacing such PHI with surrogates. Despite the recent advances in automatic de-identification of EHR, significant obstacles remain if the abundant health data available are to be used to the full potential. Accuracy in de-identification could be considered a necessary, but not sufficient condition for the use of EHR without individual patient consent. We present here a comprehensive review of the progress to date, both the impressive successes in achieving high accuracy and the significant risks and challenges that remain. To best of our knowledge, this is the first paper to present a complete picture of end-to-end automatic de-identification. We review 18 recently published automatic de-identification systems -—designed to de-identify EHR in the form of free text-— to show the advancements made in improving the overall accuracy of the system, and in identifying individual PHI. We argue that despite the improvements in accuracy there remain challenges in surrogate generation and replacements of identified PHIs, and the risks posed to patient protection and privacy.

\end{abstract}


\section{Introduction}

The application of machine learning research using EHR has the potential to revolutionise health care. There is an abundance of health data available and maximising the utility of this data will result in improving health care, especially in patient care, medical outcomes, surgical outcomes, risk prediction, clinical decision support and medical diagnosis. 

Use of patient data typically requires individual patient consent. For research, without individual consent, the data must be de-identified such that the patient's identity or privacy is not breached. Obtaining individual patient consent for massive datasets is time-consuming and is a challenging task. Hence there is a great interest in automating the de-identification process such that EHR can be used in research to improve the health care and quality of patient care without compromising the identity of the patient.   

There is growing interest internationally in applying ‘big data’ techniques to electronic health records.  However, privacy laws in many jurisdictions --including New Zealand's Health Information Privacy Code and the United States Health Insurance Portability and Accountability Act (HIPAA)-- require accurate de-identification of medical documents (such as discharge summaries and electronic health records) before they can be shared outside of their originating institutions.

The sharing of records is crucial for advancing health research. For example, the 2014 Heart Disease Risk Factors Challenge involved participating research groups attempting to predict heart disease risk factors in diabetic patients from longitudinal clinical narratives. As noted above, such a challenge would not have been possible under United States law if the narratives (1,304 medical records from 296 diabetic patients) were not de-identified first. In this case, the records were de-identified manually by multiple medical professionals. Since most institutions will not be able to afford the costs of manual de-identification, automating the process is crucial therefore for sharing data and advancing health research. 

We present our findings in three main groups: achievements, challenges, and risks, associated with generating an automatic de-identification. Achievements of automatic de-identification primarily focus on identification of PHI in EHR. Challenges are associated with the surrogate generation and replacement. Risks outline the issues relating to re-identification and medical correctness and usability of de-identified data. 

The rest of the paper is structured such that a brief background on de-identification is presented in section 2, achievements of recent de-identification systems in section 3, challenges in section 4, risks in section 5 and finally a discussion in section 6.

\section{Background on De-identification}

\begin{figure}[h!]
\begin{center}
\includegraphics[width=0.9\textwidth]{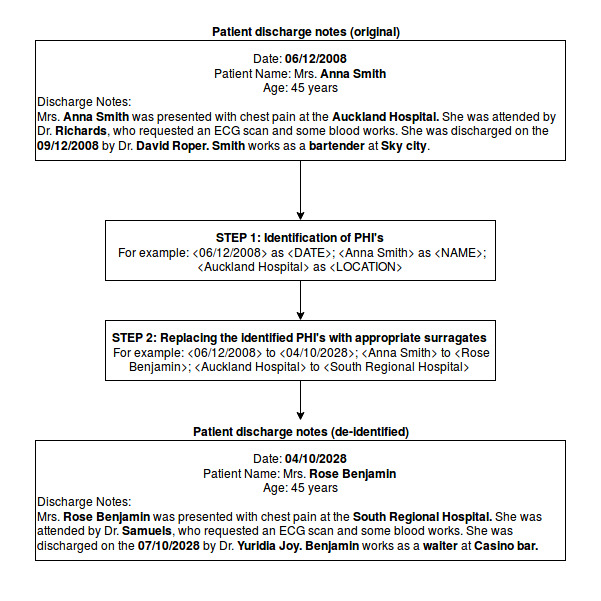}
\end{center}
\caption{Example of an end-to-end de-identification process. }
\label{fig:deidsteps}
\end{figure}

De-identification is a two-step process where PHI is identified in EHR and replaced with suitable surrogates such that patient privacy and confidentiality is not at risk. Figure \ref{fig:deidsteps} provides a detailed example of de-identification of EHR, where original patient discharge notes are de-identified.
This figure also outlines the two-step de-identification process. It is important to note that step two requires the use of appropriate surrogates to replace the original PHI and hence automating surrogate generation is a vital step in creating a longitudinal narrative end-to-end automatic de-identification system. Although EHR is in the form of tabular structures (i.e. tables), free-form narratives, and images, this study focuses on medical data in the free form longitudinal text.

De-identification should be considered a means of satisfying rather than circumventing the legal and ethical requirements created to protect patient privacy across the world. Individual countries have different requirements, for example HIPAA in the United States of America (\cite{Stubbs2015,NIST,yogarajan2018survey}), the European Union's new General Data Protection Regulation (GDPR) (\cite{brasher2018addressing,polonetsky2016shades}), and New Zealand's own health information privacy code (\cite{healthprivacy,healthdisability,yogarajan2018privacy}). HIPAA is arguably the gold standard, and both HIPAA's regulations on Expert Determination and Safe Harbor are used as the standard benchmarks for de-identification of EHR in the form of free text. We use HIPAA's Safe Harbor guidelines as the basis of assessing the accuracy of the de-identification systems (for details on HIPAA's Safe Harbor and the 18 categories see \cite{yogarajan2018survey}).

A superior de-identification system will not only meet legal requirements but will also help build societal consent by assuring the public that their privacy and medical data will be protected. This consent is vital if large-scale research involving medical records is to be accepted in the same way as, for example, Statistics New Zealand's Integrated Data Infrastructure. Acceptance of the latter is arguably in part due to measures were taken by Statistics New Zealand to de-identify data (\cite{zealand2016integrated,ragupathy2018applying}).

\section{Achievements} \label{sec:2}

In the recent years there has been a substantial development in natural language processing tasks, including de-identification, primarily due to the development in deep learning (\cite{goldberg2017neural,dalianis2018clinical}). Improving accuracy of de-identification of EHR -- step 1 from Figure \ref{fig:deidsteps} -- has been the primary focus of research in this field, and several de-identification systems have achieved remarkable success. The main reason for such development is the EHR de-identification competitions (\cite{KUMAR2015S6, naturallang, STUBBS2015S67, STUBBS2015S78, Stubbs2017, psychatric}). For a complete review of these competitions and significance see \cite{yogarajan2018survey}. It is important to note that these competitions provide open access data which allowed the research to grow rapidly. In addition to these competition datasets, the MIMIC dataset (\cite{johnson2016mimic,goldberger2000physiobank}) is another open access dataset that has been used to develop de-identification systems.

\begin{table}[h!]
\small\sf\centering
\caption{De-identification systems summary. Machine learning indicates systems that uses machine learning techniques only. Hybrid systems indicates systems that used a combination of machine learning techniques and hand crafted rules. }\label{table:reference}
\centering
\begin{tabular}{ l l }
\toprule
Architecture & De-identification system   \\
\midrule
Machine learning &S1 (\cite{zhao2018leveraging}),\\ & S2 (\cite{CHEN2015S60}) \\ & S3 (\cite{dernoncourt2017identification}), \\ &  S4 (\cite{yadav2017patient}), \\ & S5 (\cite{lee2017transfer}), \\ & S6 (\cite{DBLP:journals/corr/DernoncourtLS17}) \\
 Hybrid & S7 (\cite{YANG2015S30}) \\  & S8 (\cite{LI2017}) \\ & S9 (\cite{lee2016feature}) \\ & S10 (\cite{DEHGHAN2015S53}) \\  & S11 (\cite{YANG2015S30}) \\ & S12 (\cite{HE2015S39})  \\ & S13 (\cite{LIU2015S47}) \\ & S14 (\cite{Phuong2016AutomaticDO}) \\ & S15 (\cite{bui2017uab}) \\ & S16 (\cite{jiang2017identification}) \\ & S17 (\cite{lee2017hybrid}) \\  & S18 (\cite{kumar2016}) \\
     \bottomrule
    \end{tabular}
\end{table}

In this section, we outline the most significant achievement of automating end-to-end de-identification system: improving accuracy. It has been argued that as far as de-identification is concerned, perfection cannot be achieved; however, 95\% accuracy is considered to be the rule of thumb and universally accepted value (\cite{psychatric,STUBBS2015S11}). We use 18 de-identification systems, as outlined in Table \ref{table:reference}, to show that several of these systems have achieved an overall F-measure of $\geq$ 0.95. Also, we outline the fact that these systems have also identified the majority of the HIPAA PHI with the F-measure of $\geq$ 0.95. These achievements have been a significant milestone in automating end-to-end de-identification of EHR and has been a significant breakthrough in this area of research.

This section will be structured such that a brief overview of the datasets will be provided. This is followed by an outline of the systems that obtained an overall F-measure of $\geq$ 0.95, and also a summary of systems that recorded F-measure of $\geq$ 0.95 for individual PHIs. The techniques and datasets used for these results are also outlined.

\subsection{Overview of Datasets}
In this section, we provide a quick overview of the most commonly used datasets by the 18 de-identification systems outlined in Table \ref{table:reference}. The most commonly used datasets were introduced by the following three competitions: the 2006 Informatics for Integrating Biology and the Bedside (i2b2) competition (\cite{uzuner2007evaluating}); the 2014 i2b2/UTHealth shared task (\cite{STUBBS2015S67, STUBBS2015S78}); and the 2016 Centers of Excellence in Genomic Science (CEGS) and Neuropsychiatric Genome-Scale and RDOC Individualized Domain (N-GRID) shared task (\cite{Stubbs2017, psychatric}). 

The dataset for the 2006 competition included 889 unannotated discharge summaries, also used for smoking challenges, manually broken into sentences and tokenised. The dataset for the i2b2/UTHealth shared task 2014 included 2 - 5 records for each patient over a fixed period and was obtained from two large academic tertiary hospitals: Massachusetts General Hospital (MGH), and Brigham and Women's Hospital (BWH) \cite{KUMAR2015S7}. It includes 296 diabetics patients with 1304 longitudinal medical records and contains three cohorts based on the diagnosis of coronary artery disease (CAD) (\cite{STUBBS2015S20, STUBBS2015S67, KUMAR2015S6, STUBBS2015S78}). 

The 2016 CEGS N-GRID shared task used psychiatric data, making it the first ever competition to use psychiatric intake records (\cite{psychatric,lee2017hybrid}). The data for the 2016 competition reflected the records ``as is'' (\cite{psychatric,UZUNER2017S1}): the state at which data was received from the sources. Unlike other medical data, such as that of the 2014 challenge, psychiatric data contains an abundance of information related to the patients such as places lived, jobs held, children's ages, hobbies, traumatic events, patients' relatives' relationship information, and pet names. This makes it a much more significant challenge to de-identify (\cite{psychatric,BUI2017}).

MIMIC III  is one the most extensive publicly available database  (\cite{johnson2016mimic,goldberger2000physiobank}). It contains health records of approximately sixty thousand admissions of patients in critical care units. The database includes information such as demographics, laboratory test results, procedures, medications, and physician notes.

Also, there were other data used by individual systems such as Chinese data by S1 and Dutch data by \cite{menger2018deduce}. Although we do not describe these systems in this paper, it is important to note that these systems also presented with high F-measure.

\subsection{Overall F-measure of de-identification system}

\begin{table}[h!]
\small\sf\centering
\caption{De-identification systems with overall F-measure $\geq$ 0.95. }
\label{table:overall}
\centering
\begin{tabular}{lllll}
\toprule
   {De-identification System} &   \multicolumn{4}{c}{F-measure} \\ 
    &  i2b2 2006 & i2b2 2014  & i2b2 2016 & MIMIC III  \\
   \midrule
       S1   & 0.9879 & 0.9805 &  &    \\
        S3   & & 0.9785 &  & 0.9923   \\
       S4  & & 0.9746 & &    \\
       S5   & & 0.9800 & 0.9600  &   \\
       S6   & &  0.9770  & & \\
      S7  & & 0.9573 & &     \\ 
      S8 & & 0.9511 & &  \\
        S9  & & &  & 0.9926   \\
     \bottomrule
    \end{tabular}
\end{table}

Table \ref{table:overall} presents the de-identification systems that recorded an overall F-measure of $\geq$ 0.95. Each entry also outlines the datasets used to obtain such results. The highest recorded overall F-measure was obtained by S3 and S9 using MIMIC III dataset. One possible reason for such high accuracy obtained using MIMIC III dataset could be due to the duplicates created by cut and paste (\cite{gabriel2018presence}). The i2b2 2014 is the most commonly used dataset. It is important to point out S7 as the best performing system from the actual i2b2 2014 competition. As shown in Table \ref{table:overall} there has been a substantial improvement in
F-measure since the 2014 competition. Unfortunately, this might partly
be due to overfitting of the now known and freely available test set.

\begin{table}[h!]
\small\sf\centering
\caption{De-identification systems with overall F-measure $\geq$ 0.95. }
\label{table:overall_sys}
\centering
\begin{tabular}{ll}
\toprule
   {De-identification} &  Techniques  \\
   System & \\
    \midrule
       S1   &  Recurrent neural network (RNN) + statistical text skeleton approach. \\
        S3   &  RNN  \\
       S4  & Conditional Random Field (CRF)    \\
       S5   & Transfer Learning  \\
        S6   &  Artificial  neural  networks  (ANNs)  \\
      S7  & CRF + Rule based + Dictionary based     \\ 
      S8 & CRF + Rule based   \\
       S9  &  Long Short Term Memories (LSTMs) + human-engineered features \\
     \bottomrule
    \end{tabular}
\end{table}

Table \ref{table:overall_sys} provides an outline of the techniques used by the de-identification systems in Table \ref{table:overall}. Machine learning only systems favour deep learning approaches. Hybrid systems with the incorporation of handcrafted rules and dictionary-based approaches are also used by a couple of the de-identification systems to achieve high F-measure.

\subsection{F-measure of individual PHIs}

\begin{table}[h!]
\small\sf\centering
\caption{F-measure $\geq$ 0.95 for HIPAA categories for de-identification. On occasions where F-measure was not $\geq$ 0.95, the highest score is presented. CONTACT: URL and IP address; ID: BioID, Healthplan, Social Security no, and Vehicle licence plate no; Face photo; and Any other unique code are PHIs that were not present in any of the dataset, hence not included here.   }\label{table:HIPAA}
\centering
\begin{tabular}{ l l l l l }
\toprule
PHI categories& Sub-categories & F-measure &  Reference & Dataset \\
(HIPAA) &   & &  & \\
\midrule
{DATE} & Date & $\geq$ 0.95 &  S2, S3, S4, S8, S10 & i2b2 2014\\
& & & S11, S12, S13, S18 & \\
& & $\geq$ 0.95 & S3, S9 & MIMIC \\
& & $\geq$ 0.95 & S14 & i2b2 2006 \\
& & $\geq$ 0.95 & S15, S16, S17, S8 & i2b2 2016 \\
{NAME} & All names & $\geq$ 0.95 & S3, S18 & i2b2 2014  \\
& & $\geq$ 0.95 & S3 & MIMIC \\
{AGE} &    Age & $\geq$ 0.95 & S1, S3, S8, S18 & i2b2 2014 \\ 
& & $\geq$ 0.95 & S3 & MIMIC \\
& & $\geq$ 0.95 & S14 & i2b2 2006 \\
& & $\geq$ 0.95 & S15, S16, S17, S8 & i2b2 2016 \\
{CONTACT} &  Phone  & $\geq$ 0.95 & S3, S11 & i2b2 2014 \\
& & $\geq$ 0.95 & S9 & MIMIC \\
& & $\geq$ 0.95 & S14 & i2b2 2006 \\
& & $\geq$ 0.95 & S17 & i2b2 2016 \\
&    Fax &   0.80  & S2 & i2b2 2014 \\
&    Email & $\geq$ 0.95 & S2, S10, S11 & i2b2 2014 \\
{ID} &    Medicalrecords &  $\geq$ 0.95 & S11, S12   & i2b2 2014 \\
& & $\geq$ 0.95 & S9 & MIMIC \\
&    IDNUM & $\geq$ 0.95 & S3 & i2b2 2014 \\
&    Device & 0.80 & S2, S3 & i2b2 2014  \\
& License   & $\geq$ 0.95 & S17  & i2b2 2016   \\
{LOCATION} & all & $\geq$ 0.95 & S3  & i2b2 2014 \\ 
& & $\geq$ 0.95 & S9 & MIMIC \\
\bottomrule
\end{tabular}
\end{table}

In this section, we provide an overview of the systems that recorded F-measure $\geq$ 0.95 for individual HIPAA PHI’s. Where the F-measure was $<$ 0.95, the highest recorded score is presented. We also provide some possible issues relating to the PHI’s that have lower F-measure. This section also provides an overview of the i2b2 PHI’s. These are the additional PHI’s introduced by the i2b2 2014 and 2016 competitions  (\cite{STUBBS2015S20,STUBBS2015S67, STUBBS2015S78}). Although legally, as per HIPAA rules, there are no requirements for these additional PHI’s to be de-identified, the competition organisers argue that these extra PHIs provide more security over re-identification of data. Since i2b2 2014 and 2016 datasets are most commonly used in the advancement of de-identification research, we feel it is vital to also present the successes in these additional PHIs.  

Table \ref{table:HIPAA} provides an overview of the systems that achieved high F-measures. It also outlines the datasets that were used to obtain such results. Except for Fax and Device all other PHI’s have obtained an F-measure of $\geq$ 0.95. This is an incredible achievement and a significant improvement to the results obtained in i2b2 competitions (\cite{yogarajan2018survey}). Although Fax and Device recorded $<$ 0.95 F-measure, it is important to note that only a very few instances ($<$ 10) were found in the datasets for both of these PHIs. This makes improving the accuracy using machine learning approaches very hard.  

Table \ref{table:HIPAA_tech} provides an overview of the techniques used to obtain the F-measures presented in Table \ref{table:HIPAA}. As observed in Table \ref{table:overall_sys} there is a clear increase in deep learning methods. With a combination of handcrafted rules, de-identification systems have achieved high F-measure for the majority of the PHI. In several cases, hand-crafted rules only also achieve high F-measure. Good examples are License and Email, where regular expressions work very well.

\begin{table}[h!]
\small\sf\centering
\caption{Techniques used for the F-measures presented in Table \ref{table:HIPAA} for HIPAA PHI categories. }\label{table:HIPAA_tech}
\centering
\begin{tabular}{ l l l  }
\toprule
PHI categories& Sub-categories & Techniques \\
(HIPAA) &   &  \\
\midrule
{DATE} & Date & CRF + Rules + Dictionary (S12);  Bi-LSTM (S7);  \\ & &  CRF + Rules + Keywords (S11); \\ & &  Hidden Markov model (HMM-DP) (S2); \\ & & CRF + Rules (S10, S14, S15, S17); LSTM (S16); \\ & & RNN (S3, S18); LSTM + Rules (S9); CRF (S4)  \\
{NAME} & All names & RNN (S3, S18)    \\
{AGE} &    Age & Bi-LSTM (S7); CRF +  Rules (S14, S15, S17);\\  & & LSTM (S16); RNN (S1, S3, S18)  \\ 
{CONTACT} &  Phone  & CRF + Rules +  Keywords (S11); RNN (S3);  \\ & & CRF + Rules (S14, S17); LSTM + Rules (S9)  \\
&    Fax &  HMM-DP (S2)  \\
&    Email & Rules (S10, S11); HMM-DP (S2) \\
{ID} &    Medicalrecords & CRF + Rules + Keywords (S11); LSTM + Rules (S9)  \\
&    IDNUM & RNN (S3)  \\
&    Device & HMM-DP (S2); RNN (S3)  \\
& License   & Rules (S17) \\
{LOCATION} & all & RNN (S3); LSTM + Rules (S9)  \\ 
\bottomrule
\end{tabular}
\end{table}

Table \ref{table:i2b2extra} provides an overview of F-measures for i2b2 introduced extra PHIs. These PHIs are not part of the legal requirement as per HIPAA regulations but are additional security for ensuring that patient privacy and confidentiality are maintained. Compared to the recorded results in the i2b2 2014 and 2016 competitions, there is a substantial increase in the F-measure. Clearly, Organisation, Location-others, Profession and Country are the PHI’s yet to reach the 0.95 F-measure. These were also the PHI’s that recorded a very low F-measure in both competitions (see \cite{yogarajan2018survey} for details). The main issue with Country and Organisation is that the data is very sparse. Location-others only occurs in thirteen instances in the dataset. The sparsity of the data and the very low frequencies of same values make achieving higher F-measures very hard. However, there is still an improvement in results compared to that recorded in the competitions.

\begin{table}[h!]
\small\sf\centering
\caption{The best F-measure for i2b2 extra categories for de-identification. This table includes categories not included in Table \ref{table:HIPAA}, but were introduced by i2b2 competitions as additional categories (\cite{STUBBS2015S11,STUBBS2015S20}). It also provides the techniques used to achieve these F-measures. }\label{table:i2b2extra}
\centering
\begin{tabular}{ l l l l l l }
\toprule
PHI categories& Sub-categories& F-measure &  Reference & Techniques & Dataset \\
(i2b2 extra) &  & &  & \\
\midrule
{NAME} & Doctor & $\geq$ 0.95 & S3, S4  & RNN; CRF & i2b2 2014 \\
& &$\geq$ 0.95 & S2 & LSTM + Rules & MIMIC \\
& & $\geq$ 0.95 & S14 & CRF + Rules & i2b2 2006 \\
& & $\geq$ 0.95 & S17 & CRF + Rules  & i2b2 2016 \\
&    Patient & $\geq$ 0.95 & S3, S4 & RNN; CRF & i2b2 2014 \\
& &$\geq$ 0.95 & S2 & LSTM + Rules & MIMIC \\
& & $\geq$ 0.95 & S14 & CRF + Rules & i2b2 2006 \\
&    Username &  $\geq$ 0.95 & S10 & Rules; & i2b2 2014  \\
& &$\geq$ 0.95 & S11 & CRF + Rules + KW; & i2b2 2014  \\
& &$\geq$ 0.95 & S12 & CRF + Rules + Dic & i2b2 2014  \\
{LOCATION} & Hospital & $\geq$ 0.95 & S3 & RNN & i2b2 2014 \\
& & $\geq$ 0.95 & S2 & LSTM + Rules & MIMIC \\
&    City & $\geq$ 0.95 & S3 & RNN & i2b2 2014 \\
&    State & $\geq$ 0.95 & S3 & RNN & i2b2 2014 \\
& & $\geq$ 0.95 & S2 & LSTM + Rules & MIMIC \\
&    Street & $\geq$ 0.95 & S3, S9 & RNN; HMM-DP & i2b2 2014 \\
& &$\geq$ 0.95 & S11 & CRF + Rules + KW; & i2b2 2014  \\
&    Zip & $\geq$ 0.95 & S10& Rules & i2b2 2014 \\
& &$\geq$ 0.95 & S11 & CRF + Rules + KW; & i2b2 2014 \\
& & $\geq$ 0.95& S12 & CRF + Rules + Dic & i2b2 2014  \\
& & $\geq$ 0.95 & S2 & LSTM + Rules & MIMIC \\
& & $\geq$ 0.95 & S16 & LSTM & i2b2 2016 \\
&    Organisation & 0.82 & S3 & RNN & i2b2 2014  \\
&    Country & $\geq$ 0.90   & S3 & RNN & i2b2 2014  \\
& & $\geq$ 0.90  & S2 & LSTM + Rules & MIMIC \\
&    Location-Others & 0.57 & S3 & RNN & i2b2 2014 \\
{PROFESSION} &  Profession & 0.84 & S3 & RNN & i2b2 2014 \\
\bottomrule
\end{tabular}
\end{table}

\subsection{In summary}
This section showed the achievements in automating de-identification research, with substantial improvement in F-measure of identifying PHI in the overall systems and individual PHI’s (notably the HIPAA required PHI’s).

\section{Challenges}

The biggest challenge in automating end-to-end de-identification is surrogate generation and surrogate replacement (step 2 in Figure \ref{fig:deidsteps}). At first sight, this appears to be superficially simple when compared to step 1. However, when one considers it in detail, there are many complex subtleties associated with the surrogate generation and surrogate replacement. Unlike the research towards increasing accuracy in identifying PHI, as seen in section \ref{sec:2}, this is an area where very little research progress has been made. There have been only a few papers published in the recent years regarding surrogate generation and surrogate replacement for the de-identification problem, with the schema developed in the 2014 i2b2 competitions being the prominent one to date (\cite{stubbs2015challenges,stubbs2015annotating}). 

PHI are categorised into explicit identifiers and quasi-identifiers. Explicit identifiers such as name, phone number and social security number are directly linked to a patient. Quasi-identifiers such as age, gender, race and zip code are not directly connected to a patient but can be linked to external data sources and consequently be used to identify a patient, hence posing the same risk to patient privacy as explicit identifiers.

In this section, we present examples of standard practices used in surrogate replacement and challenges faced. Automation in the surrogate generation is arguably still a very challenging and unsolved problem. 

\subsection{Examples of Surrogate replacement of PHI}

Table \ref{table:surrogate} provides an outline of some PHIs and the standard practices used in a surrogate generation while creating de-identified data. Although all of these practices are based on hand-crafted rules and pre-compiled tables, there was also a need to do a manual check after the data is de-identified. This is to ensure that medical correctness, readability and consistency are maintained
across the health data. Table \ref{table:surrogate} also indicates where manual checking after de-identification was required. Surrogates need to maintain the same form as the original, and where possible same internal temporal and co-reference relationships. Also, as illustrated in Figure \ref{fig:deidsteps} semantic links must be maintained, for example between LOCATION and PROFESSION. 

\begin{table}[h!]
\small\sf\centering
\caption{Common practices used in surrogate generation and replacement of PHI as outlined in (\cite{stubbs2015challenges,stubbs2015annotating,pantazos2017preserving,johnson2016mimic}). }\label{table:surrogate}
\centering
\begin{tabular}{ l l l}
\toprule
PHI & Surrogate generation techniques & Manual check \\
\midrule
 DATE and AGE & Option 1: date shifting where all elements & Yes \\  & of dates (i.e. day, month and year) are shifted \\ & forward by the same random number. &\\ 
&   Option 2:  distorted identifier table is used where & \\ & date, month were changed but year was kept the same. & \\  & & \\
PHONE, FAX,  & Randomly created surrogates.& - \\ URLs, ID  & & \\ & & \\
EMAIL address & Manual replacement & Yes\\ & & \\
NAME & Option 1: permutation tables with existing  & - \\ & identifiers are mapped to new ones with & \\ & similar frequency of occurrence. & \\
& Option 2: Mapping between letters, maintaining & \\ & name type and sex. & \\ & & \\
LOCATION & Random selection of surrogates from pre-compiled &  Yes \\ &  list or permutation tables ensuring the type of & \\ &  location is matched. & \\ & & \\
PROFESSION & hand-crafted rules to select from pre-compiled list. & Yes \\ & & \\ 
\bottomrule
\end{tabular}
\end{table}

It is important to note that it is relatively easy to create surrogates randomly and maintain co-references for PHONE, FAX, URLs and ID (\cite{stubbs2015challenges,stubbs2015annotating}). Any ambiguous words appearing as part of a name, medical term or acronym were replaced using a set of hand-crafted rules (\cite{pantazos2017preserving}). This is primarily because, in medicine, it is common to have diseases, signs and symptoms being named after the person first describing it. One such example is ``Aaron'' which can refer to a name of a person, or be part of a medical term: Aaron sign referring the pain felt in the epigastrium.

\subsection{Issues and Challenges due to Surrogate replacement of PHI}

Table \ref{table:surrogate} provided an overview of techniques used in the surrogate generation and replacement. However, there are many practical issues with some of these rules and techniques which creates challenges in maintaining medical correctness and usability of de-identified data in health advancement research. Moreover, it is important to note in most cases there was a need to manually check the surrogate replaced data to ensure consistency and accuracy is maintained across patient data.  

When DATE is changed to just the year or randomly changed it removes inferrable information such as the ``season'' which could result in missing any pandemic outbreak (\cite{li2017anonymizing}). There is a need to maintain the semantic link between the LOCATION and DATE to ensure such information is not missed. Also, for PHIs DATE and AGE, medical correctness is a major issue. Birth dates have to be transformed such that the patient age is in a similar age range. Otherwise diagnosis patterns will become inapplicable. For example, a 20-year-old de-identified to be a 60-year-old will cause issues in medical diagnosis.

When LOCATION such as zip code is replaced by random zip codes (even from a pre-compiled list), geographical information is distorted. For example, a patient living in a high socioeconomic area being moved to low decile area, or vice versa, will result in relevant information about the living conditions and life expectancy changing. This could mislead patient diagnosis, or miss vital information in patient care. In addition to socioeconomic issues relating to LOCATION, there is also ethnicity information. For example, in New Zealand, there are parts of the country, such as Northland, where there is known to be a higher population of New Zealand's indigenous Maori people. If everyone from Northland is moved to another LOCATION or spread across several LOCATIONS, the ethnicity information is also lost in the de-identified data. It is very challenging to ensure such information is not lost without introducing systemic bias towards a sub-population, e.g. Maori people in the New Zealand example above.

With NAME, if the patient's name, for example, ``John'', is replaced by ``Jack'', then there is a need to ensure all of his medical records reflect this change. For instance, in addition to the free text data that was replaced, the change must also be made consistently across all of his longitudinal data, including but not restricted to his structured data and medical images. In addition, the name change should also reflect correctly on his family's medical records, i.e. his wife's records and his children’s. This does allow the consistency and medical correctness to be maintained in de-identified data (\cite{pantazos2017preserving}). The need to maintain consistency and medical correctness makes automating de-identification a very challenging task and does require manual checks and inputs (\cite{stubbs2015challenges,pantazos2017preserving}). Also, in order to maintain readability, a patient name must be replaced by a new name that looks real and consistency should also be taken into consideration. For example, the frequency of the name in a database needs to be consistent. A rare name occurring more frequently will not look real.

One of the many challenges faced in de-identifying medical, free text data is ambiguous words. In many cases, it is challenging to differentiate between an everyday word, medical word and part of the patient name. This may result in errors with surrogate replacement where for instance a medical term is replaced by a person's name surrogate. 

\subsection{In summary}
This section presented common practices used in surrogate generation and replacement, where most of the techniques rely on hand-crafted rules and pre-compiled tables. We outline some of the important challenges faced in this step of de-identification and argue that automation in surrogacy is still an open question with many obstacles to overcome.

\section{Risks}

De-identified data in addition to protecting patient privacy should also meet the following standards: medical correctness, readability and consistency across data (\cite{pantazos2017preserving}). Risks around de-identification of health data can be classified into two main areas: the risk of re-identification and the risk of losing usability, medical correctness and consistency across data. In this section, we provide a brief overview of these two areas. 

\subsection{Re-identification}

Re-identification is a process where a person's identity is identified from the de-identified data. This does result in a serious breach of patient privacy and confidentiality. Explicit identifiers such as person's name and address can be considered obvious identifiers. However, even quasi-identifiers can result in re-identification of a person (\cite{johnson2018evaluation,sweeney2002k,li2017anonymizing}). There have been many examples of such occurrences where quasi-identifiers have been matched with external resources to identify patients. For example, it was proven that attributes such as gender, date of birth and zip code could be matched with external sources such as voting data to identify a patient (\cite{sweeney2002k,li2017anonymizing}). Also, other examples demonstrate that a combination of a small subset of quasi-identifiers, with or without other medical data, may even be enough to identify the individual patient and pose serious threat to patient privacy (\cite{el2006evaluating,Mayo2019}).   

In addition to explicit identifiers and quasi-identifiers, there are also the sensitive attributes such as psychiatric diseases, HIV, and cancer, which patients are not willing to be associated with. Due to the specific nature of these sensitive attributes and the need for special care facility these attributes when combined with other identifiers makes re-identification of a patient much more feasible (\cite{gkoulalas2014publishing}).   

The risk of re-identification is real and can lead to serious breaches
of patient privacy and confidentiality. While designing an automatic de-identification system, it is important to consider the re-identification risk and take appropriate measures to minimise such risk. Also, there needs to be transparency in acknowledging such concerns. The main questions when it comes to re-identification are:
\begin{itemize}
    \item what is the accepted level of risk with re-identification?
    \item who makes that decision, the de-identification system designers, the users or the patients? 
\end{itemize}
 There is no easy or correct answer to these questions, but they still need to be considered when designing a de-identification system. There is a need for human input in making such decisions and deciding the boundaries of acceptable risk associated with de-identification of a medical system.

\subsection{Medical Correctness and Usability of De-identified Data}

Maintaining medical correctness, consistency, readability and usability of data is a difficult problem and the risks associated with this are usually overlooked. Compared to de-identification of structured data, unstructured free text is very challenging. It contains medical information about a  patient that needs to be preserved for medical correctness. However, it also contains personal details such as name, phone number, family members names and other personal identifying items. Although the accuracy of identifying PHI in such data has improved considerably, ensuring these PHI are replaced with appropriate surrogates, and medical correctness maintained, is an open question. This poses a great risk in using de-identified data for machine learning based health research, as the de-identified health records may compromise the accuracy and outcome of the resulting model. For example, accidentally replacing a word that resembles a name but is not a name (maybe an abbreviation for a disease, or a disease name itself) can result in readability and medical correctness errors (\cite{pantazos2017preserving}). The hope is that the original data and the de-identified data of a particular problem will result in the same outcome. However, there is no clear evidence that it does. In reality, the only way to check if it does or does not, is by
building models for both versions of the data and comparing them.

Many of the surrogate replacements use randomised identifiers. However, in such cases, the readability and consistency are compromised (\cite{pantazos2017preserving}). Unless manually checked there is no guarantee these randomly replaced PHI makes much sense in the context and provide useful data outcomes.

Another significant risk is accidentally confusing patients. Let’s say you have two patients in the same age range, both named Anne Smith, but one presenting with cancer and the other one with
cardiovascular issues. Ensuring these two are kept separated across all of their data, especially longitudinal data, can be very hard. This will require using several PHI to match the person's identity. However, in this case, there is an increased risk of re-identification. This poses a question around confidentiality vs verifiability, and as a result, increases the risk. This problem cannot be readily solved by using unique identifiers (such as NHI numbers, date of birth or tax numbers) to match narratives, as automated de-identification systems by design prune such de-identifiers. Furthermore, HIPPA’s Safe Harbor provision mandates the removal of such unique identifiers (\cite{Stubbs2015,NIST}). Similarly, New Zealand’s Privacy Act and Health Information Privacy Code set strict limits on the assignment and use of unique identifiers (\cite{healthprivacy,healthdisability}).

\subsection{In Summary}

This area outlined the two main risks associated with de-identification: the risk of re-identification and the risk of losing usability, medical correctness and consistency across data. Minimising the risk posed to patient privacy and confidentiality is vital. The risk of re-identification must be considered a severe threat when designing a de-identification system. The de-identified data must also maintain medical correctness, readability and consistency. The advancement of health research using de-identified data does rely on the usability of data and medical correctness of data. There is a need for a manual check to ensure that the de-identified data resembles the original data.  

\section{Discussion}

To best of our knowledge, this is the first paper to present a complete picture of end-to-end automatic de-identification of medical narratives. Noticeably, the majority of the research is done on improving the accuracy of PHI identification in the overall system and of individual PHI. We acknowledge the need for such research, and despite the recent advancements in this area, mainly due to the use of deep learning in natural language processing tasks, there is more room for improvement. At this stage the minimum requirement of 95\% F-measure has been met by several systems, but this is only the minimum requirement. There is room for higher F-measures. Also, it will be nice to take these systems to the next level, where in addition to the open access data they use other sources of data. It will be interesting to see the adaptability of these systems.

One of the big downfalls to these systems is that they do not outline the surrogacy generation aspect of de-identification. However, de-identification is not just about identifying the PHI, but is also replacing the identified PHI with appropriate surrogates. As mentioned earlier there is very little research done in this area, and clearly, there are many challenges yet to be overcome. Also, most of the current practices are data specific and use hand-made rules and pre-compiled tables. This is far from achieving full automation in the de-identification problem, and there is an explicit acknowledgement of the need for manual checks after surrogate replacement. We encourage for more research in this area, where the priority is to address some of the challenges outlined in this paper and also to eliminate the need for manual checks.  

The importance of automatic de-identification in advancing health research cannot be emphasised enough. However, there is still a need to be aware of the risks associated with designing such systems. This will ensure that risk to patient privacy and confidentiality is minimised while advancing the field of medicine through maximising the potential of EHR with the use of machine learning techniques. Also, it is vital that the medical correctness, consistency, readability and usability of data are all maintained such that the resulting de-identified data provides the parallel output to that of the original data. It must be pointed out that high accuracy of de-identification is directly proportional to medical correctness. It does become harder to maintain the medical correctness and usability of data when achieving high accuracy becomes the focus. Hence, de-identification of data does become a balancing act where barriers associated with risk and benefits must both be considered. This is another area where there needs to be more research done in proving that de-identified data is providing the same outcomes as the original data.

The challenges and risks associated with de-identification have opened new avenues of research in finding alternatives. One has to ask the question: ``What if proper de-identification is impossible?''. \cite{data2018alternative} proposes an alternative idea for sharing confidential data called ``model to data'' where the flow of information between data generators and modellers is reversed. Another idea presented by \cite{vepakomma2018split} proposes a deep learning model which excludes the need to share raw patient data or labels. These are merely examples of alternatives, and are just the beginning of possibly solving the problem of sharing and using EHR without the risk to patient privacy and confidentiality.

\bibliographystyle{chicago}
\bibliography{bibliography_deid}

\end{document}